\documentclass[preprint,12pt]{elsarticle}

\usepackage{graphicx}
\usepackage{amsmath,amssymb}
\usepackage{array}
\usepackage{color}
\usepackage{amsbsy}
\renewcommand\vec[1]{\ensuremath\boldsymbol{#1}}

\newcommand{\h}{\mathrm{\textbf{H}}}
\newtheorem{definition}{Definition}

\newcommand{\ket}[1]{| #1 \rangle}

        \def\cH{{\cal H}}

\def\cP{{\cal P}}        

        \def\cT{{\cal T}}

\def\cX{{\cal X}}

\begin{document}

\begin{frontmatter}

\title{Multicasting Homogeneous and Heterogeneous Quantum States in Quantum Networks}


\author[MIT]{Yi-Chang Shih\fnref{was}}
\ead{yichang@mit.edu}
\author[Hsieh]{Min-Hsiu Hsieh}
\ead{minhsiuh@gmail.com}
\author[NTU]{Hung-Yu Wei\corref{cor1}}
\ead{hywei@cc.ee.ntu.edu.tw}

\fntext[was]{This research was done while the author was with National Taiwan University}

\cortext[cor1]{Corresponding author}

\address[MIT]{ Department of Electrical Engineering and Computer Science, Massachusetts Institute of Technology. }
\address[Hsieh]{Statistical Laboratory, University of Cambridge, \\
Wilberforce Road, Cambridge CB3 0WB, UK.}
\address[NTU]{Department of Electrical Engineering and Graduate Institute of Communication Engineering, National Taiwan University.}

\begin{abstract}
In this paper, we target the practical implementation issues of quantum
multicast networks. First, we design a recursive lossless compression that
allows us to control the trade-off between the circuit complexity and the
dimension of the compressed quantum state. We give a formula that describes the
trade-off, and further analyze how the formula is affected by the
controlling parameter of the recursive procedure. Our recursive lossless
compression can be applied in a quantum multicast network where the source
outputs homogeneous quantum states (many copies of a quantum state) to a set of destinations through a bottleneck.
Such a recursive lossless compression is extremely useful in the current
situation where the technology of producing large-scale quantum circuits is
limited. Second, we develop two lossless compression schemes that work for
heterogeneous quantum states (many copies of a set of quantum states) when the
set of quantum states satisfies a certain structure. The heterogeneous
compression schemes provide extra compressing power over the homogeneous
compression scheme. Finally, we realize our heterogeneous compression schemes in
several quantum multicast networks, including the single-source multi-terminal
model, the multi-source multi-terminal model, and the ring networks. We then 
analyze the bandwidth requirements for these network models.
\end{abstract}

\end{frontmatter}

\section{Introduction}
Multicasting a number of information sources to a set of destinations in a
classical network can be efficiently performed if each node is allowed to
employ additional encoding operations \cite{Ahlswede00networkinformation}.
Mixing, compressing, or distributing data at intermediate network nodes is
generally referred to as ``network coding'' \cite{Ho06arandom}.
Store-and-forward routing technique had been the dominant mainstream for
transmitting classical information through a network until network coding was invented.
In contrast to the intuitive way to operate a network that tries to avoid
collisions of data, classical network coding provides a plethora of surprising
results and opens up many practical applications in information and coding
theory, networking, switching, wireless communications, cryptography, computer
science, operations research, and matrix theory \cite{citeulike:1181440}.

An easily neglected but critical operation in classical networking coding is
the ability to clone or, simply put, to copy classical data. In the simplest
example of the butterfly network \cite{Ahlswede00networkinformation}, network
coding enables two senders to transmit one bit, respectively, to two receivers
only if the nodes of the network can make copies of the classical data
\cite{KGNR09}. Copying classical data is so straightforward that we seldom
stress its importance in a protocol until we encounter the no-cloning theorem
\cite{noclone} in the quantum domain. This theorem prohibits copying an unknown
pure quantum state, and is generalized to include the mixed quantum states that
are noncommuting \cite{nobroadcast}.

The no-cloning theorem posts a strict limitation on what can be done in a
quantum network. Hayashi \emph{et al.} showed that sending two qubits
simultaneously and perfectly in the butterfly network is impossible
\cite{HINRY07}. Leung, Oppenheim and Winter extended this impossibility result
to classes of networks other than the butterfly network \cite{butterfly}.
Faithfully transmitting quantum states in a quantum network can be achieved
when extra resources are available or specific assumptions are made for the
quantum networks. Hayashi constructed a protocol that transmits two quantum
states perfectly in the butterfly network when prior entanglement shared
between two senders is available \cite{Hayashi07}. Kobayashi \emph{et al.}
showed that perfect quantum network coding is achievable whenever classical
network coding exists if two-way classical communication is available
\cite{KGNR09}. On the other hand, if the quantum data to be multicast are
composed of the same quantum states, Shi and Soljanin proposed a lossless
compression scheme as a mean to implement quantum network coding. Their method
achieves simultaneous and perfect transmission, and the bandwidth (edge
capacity) in quantum multicast networks can be significantly reduced
\cite{multicast}.

In this paper, we design an efficient and recursive implementation of the
lossless compression proposed in \cite{multicast} for quantum multicast
networks. The implementation complexity of the network coding scheme in
\cite{multicast} grows exponentially with the number of receivers, $N$. Our
recursive compression procedure provides a trade-off between the hardware
complexity and the required bandwidth. We further propose two lossless
compression schemes that improve the compressing power of \cite{multicast} when
the set of quantum states satisfies a certain structure. We apply the two
lossless compression schemes in several quantum multicast networks, and analyze
the bandwidth reductions in quantum multicast networks.

There are many other related works on distributed computation, secret key
sharing, and key distribution over quantum networks. Van Meter, Nemoto, and
Munro investigated and analyzed applications of quantum error correction codes
on distributed computation over quantum networks \cite{distributed}. Cheng,
Wang, and Tao designed a quantum communication protocol for wireless networks
based on quantum routing \cite{wireless}. Ma and Chen suggested using the GHZ
states for multiparty secret sharing over quantum networks
\cite{multipartyGHZ}. Deng \emph{et al.} considered using EPR pairs for sharing
a quantum state among many parties in a quantum network \cite{multipartyEPR}.

This paper is organized as follows. We review the necessary materials in order
for the readers to understand the rest of the paper in section~\ref{sec_II}. In
section \ref{secthomo}, we introduce our recursive lossless compression
procedure, and derive a formula that describes the trade-off between the
circuit complexity and the dimension of the compressed state. The recursive
lossless compression can be employed to multicast homogeneous quantum states in
quantum networks. In section \ref{secthetero}, we propose two lossless
compression schemes for heterogeneous quantum states when the set of
quantum states possesses a certain structure. We compare and analyze the
homogeneous and heterogeneous encoding methods in quantum multicast networks,
including the multi-source
multi-terminal model and the ring network in section~\ref{sectappli}. We
conclude in section \ref{sectdiscuss}.

\section{Preliminaries}\label{sec_II}
A pure quantum state $\ket{\phi}$ can be mathematically represented as a column vector with unit length in a $d$-dimensional Hilbert space $\cH_d$ that is spanned
by an orthonormal basis $\{\ket{i}\}_{i=0}^{d-1}$. When $d=2$, quantum states
are called qubits (qudits for $d\geq 3$). A quantum system that contains $N$
identical copies of a quantum state $\ket{\phi}$ is denoted as
$\ket{\phi}^{\otimes N}$. For example, when $N=d=2$, the state
$\ket{\phi}^{\otimes 2}$ is
$$|\phi\rangle^{\otimes
2}=\alpha^2|00\rangle+\alpha\beta(|01\rangle+|01\rangle)+\beta^2|11\rangle,$$
where we denote $\ket{\phi}\equiv \alpha\ket{0}+\beta\ket{1}$
($|\alpha|^2+|\beta|^2=1$). A general two-qubit quantum state $\ket{\psi}$ can
be described as the following unit vector in a 4-dimensional Hilbert space
$\cH_2^{\otimes 2}$:
\begin{equation}\label{psi1}
\ket{\psi}=\alpha\ket{00}+\beta\ket{01}+\gamma\ket{10}+\delta\ket{11}.
\end{equation}
The state $\ket{\psi}$ in (\ref{psi1}) is \emph{entangled} if and only if it
cannot be written in terms of the tensor product of two single-qubit states
\cite{NC00}. One such example is when we choose
$\alpha=\delta=\frac{1}{\sqrt{2}}$ and $\beta=\gamma=0$, $\ket{\psi}$
corresponds to the maximally entangled state.

Let the set $\cX=\{0,1,\cdots,d-1\}$. Denote by $\cX^N$ the set of all possible
sequences $x^N=(x_1,x_2,\cdots,x_N)$, where each $x_i\in\cX$. Denote by
$t(a|x^N)$ the number of occurrences of the symbol $a\in\cX$ in the sequence
$x^N$. The \emph{type} of a given sequence $x^N$ is the empirical distribution:
$$P_{x^N}(a)=\frac{t(a|x^N)}{N}, \ \ \forall a\in\cX.$$
Denote by $\cT_P$ the collection of sequences that give the same empirical
distribution $P=(P_0,P_1,\cdots,P_{d-1})$:
$$\cT_P=\{x^N: P_{x^N}(a)=P_a, \ \ \forall a\in\cX\}.$$
Denote by $\cP_N(\cX)$ the set of all possible types in $\cX^N$. We have
\cite{CT91}:
\begin{equation}
|\cP_N(\cX)|=\h^N_d \leq (N+1)^d,
\end{equation}
where the function $\h^N_d$ is given by
$$\h^N_d=C^{N+d-1}_d=\frac{(N+d-1)!}{d!(N-1)!}.$$

We can construct a permutation-invariant basis
$\{\ket{\vec{e}_P}\}_{P\in\cP_N(\cX)}$ for the Hilbert space $\cH_d^{\otimes
N}$:
\begin{equation}\label{eq_PI}
\ket{\vec{e}_P}\equiv \frac{1}{\sqrt{|\cT_P|}}\sum_{x^N\in\cT_P}\ket{x^N}.
\end{equation}
Any quantum state $\ket{\Phi}\equiv\ket{\phi}^{\otimes N}$, where
$\ket{\phi}=\sum_{x=0}^{d-1}p(x)\ket{x}$, can be expressed in terms of
$\{\ket{\vec{e}_P}\}_{P\in\cP_N(\cX)}$:
\begin{align*}
\ket{\Phi}&=\sum_{x^N\in\cX^N}p^N(x^N)\ket{x^N} \\
&=\sum_{P\in\cP_N(\cX)} q_P  \ket{\vec{e}_P}
\end{align*}
where $p^N(x^N)\equiv p(x_1)p(x_2)\cdots p(x_N)$ and
$$q_P=\sqrt{|\cT_P|}\prod_{a=0}^{d-1}p(a)^{NP_a}.$$
We can define a one-to-one mapping that maps each $P\in\cP_N(\cX)$ to a number $s_P$ in
$\{0,1,\cdots,\h_d^N-1\}$ since the size of $\cP_N(\cX)$ is $\h_d^N$. 
There exists a unitary transformation $U$ such that,
$\forall P\in\cP_N(\cX)$,
\begin{eqnarray}\label{singleencode1}
U|\vec{e}_P\rangle = |0\rangle^{\otimes \left(N-\log_d \h^N_{d}\right)}\otimes|s_P\rangle,
\end{eqnarray}
where $\{\ket{s_P}\}$ forms an orthonormal basis for $\cH_d^{\otimes (\log_d \h^N_{d})}$.
Then
\begin{eqnarray}\label{singleencode}
U|\Phi\rangle = |0\rangle^{\otimes \left(N-\log_d \h^N_{d}\right)}\otimes\left(\sum_{P\in\cP_N(\cX)}q_P|s_P\rangle\right).
\end{eqnarray}
Denote $\ket{\Phi'}\equiv\sum_{P\in\cP_N(\cX)}q_P|s_P\rangle$. The dimension of
$\ket{\Phi'}$ is $\h^N_d$.

Let $A$ be the operation that adds the ancilla state $\ket{0}$ to a quantum
state $\ket{\Psi}$:
$$A:\ket{\Psi}\to \ket{0}\otimes\ket{\Psi}.$$
Let $R$ be the operation that removes the ancilla state $\ket{0}$ from a
quantum state:
$$R:\ket{0}\otimes\ket{\Psi} \to \ket{\Psi}.$$
We further assume that the operations $A$ and $R$ can add or remove as many
ancilla states $\ket{0}$ as we need in the protocol.

The unitary $U$ in (\ref{singleencode1}) followed by the operation $R$
implements a lossless compression that compresses the original state
$\ket{\Phi}$ of dimension $d^N$ to the state $\ket{\Phi'}$ of dimension
$\h_d^N$:
\begin{equation}\label{compress1}
RU\ket{\Phi}=\sum_{P\in\cP_N(\cX)}q_P|s_P\rangle=\ket{\Phi'}.
\end{equation}
Furthermore, the compression is lossless because we can recover the original
state $\ket{\Phi}$ from $\ket{\Phi'}$:
\begin{equation}\label{decompress1}
U^{-1}A\ket{\Phi'}=\ket{\Phi}.
\end{equation}
The main reason that (\ref{compress1}) and (\ref{decompress1}) hold is because
the first $(N-\log_d \h_d^N)$ qudits in (\ref{singleencode}) are in the ancilla
states and are not entangled with the last $\log_d \h_d^N$ qudits.

\section{Recursive homogeneous encoding in quantum multicast networks}\label{secthomo}

\begin{figure}
\centering
\includegraphics[width=7cm , height = 5 cm]{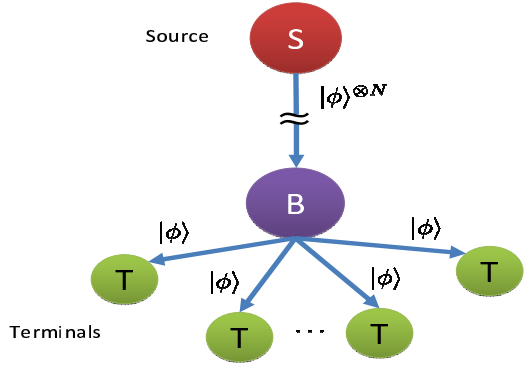}
\caption{(Color online) Quantum multicast networks that contain a single source
(labeled as $S$) and $N$ terminals (labeled as $T$). The source $S$ can generate
$N$ copies of an identical quantum state $\ket{\phi}$. The multicasting task is
for the source $S$ to distribute the quantum state $\ket{\phi}$ to each terminal $T$ simultaneously
and perfectly through the link that connects the source $S$ and the node $B$.}
\label{singledata}
\end{figure}

The main result of this section is a recursive encoding for the quantum
multicast network depicted in Fig.~\ref{singledata}. The network contains a
single source and $N$ terminals. The source $S$ can generate $N$ copies of an
identical quantum state $\ket{\phi}$ in $\cH_d$. The multicasting task is for
$S$ to distribute the quantum state $\ket{\phi}$ to each terminal $T$
simultaneously and perfectly through the link that connects the source $S$ and
the node $B$. If no encoding is performed at the source $S$, the bandwidth
(edge capacity) between the source $S$ and the node $B$ must be at least $N$
qudits per channel use in order for the source to transmit $\ket{\phi}^{\otimes
N}$ faithfully. Shi and Soljanin showed that there exists an encoding that can
reduce the bandwidth from $N$ qudits per channel use to $\log_d \h_{d}^N$
qudits per channel use \cite{multicast}. The encoding operation in
\cite{multicast} corresponds to a $d^{N}$ by $d^{N}$ unitary matrix, and is
reviewed in (\ref{singleencode1}). In the following, the encoding unitary $U$
acting on $N$ copies of a $d$-dimensional quantum state $\ket{\phi}$ is denoted
by $U^N_d$. The complexity of the encoding circuit that implements this matrix
$U^N_d$ increases exponentially with the number of terminals $N$, and quickly
becomes unfeasible even for small $N$.

We propose a recursive encoding circuit that
can provide a trade-off between the hardware complexity and the bandwidth consumption.
The idea is, instead of encoding $N$ copies of the quantum state $\ket{\phi}$
as a whole, to divide the quantum data $\ket{\phi}^{\otimes N}$ into
$\frac{N}{k}$ groups where each group contains $k$ copies of the state
$\ket{\phi}$. We then encode each of these groups to a quantum state, say
$\ket{\phi'}$, by a smaller encoding unitary $U^k_d$ of size $d^k\times d^k$.
The dimension of the compressed state $\ket{\phi'}$ is $\h^k_d$. Finally, we
encode $\frac{N}{k}$ copies of the state $\ket{\phi'}$ by another encoding
unitary matrix $U^{N/k}_{\h^k_d}$. The above 2-step encoding process can be
generalized to the following recursive encoding:
\begin{enumerate}
  \item If $N=1$, abort. Otherwise, divide $N$ copies of the quantum state
      $\ket{\phi}$ into smaller groups, where each group contains $k$
      copies of the state $\ket{\phi}$. Denote $\ket{\Phi}\equiv
      \ket{\phi}^{\otimes k}$.
  \item Encode the quantum state $\ket{\Phi}$ in each group by the encoding
      unitary $U^k_d$ .
  \item Throw away the first $(k-\log_d \h^k_{d})$ redundant qudits. Denote
      by $\ket{\Phi'}$ the rest $(\log_d \h^k_d)$-qudit quantum state.
  \item Set $N \leftarrow \frac{N}{k}$, $d \leftarrow \h^k_d$, and
      $\ket{\phi} \leftarrow \ket{\Phi'}$. Go to step 1.
\end{enumerate}
We refer to this process as the \textit{recursive homogeneous encoding}.

Denote by $y_n$ the dimension of the compressed quantum state $\ket{\Phi'}$ at
$n^{th}$ step of the above recursive procedure. Let $y_0= d$, which corresponds
to the dimension of the original quantum state $\ket{\phi}$. We then have the
following recursive relation:
\begin{eqnarray}\label{recurexact}
y_{n+1} = \h^{k}_{y_n}.
\end{eqnarray}
Let $L_n$ denote the number of the remaining qudits  after the $n^{th}$
recursion:
\begin{eqnarray}\label{def_Ln}
L_n=\log_d y_n.
\end{eqnarray}
In our recursive protocol, there are $n^*\equiv\log_k N$ recursions. Denote by
$L=L_{n^*}$ the number of remaining qudits after the last encoding unitary. The
parameter $L$ describes the minimal bandwidth requirement for the link
connecting the source $S$ and the node $B$ in Fig.~\ref{singledata} when the
source $S$ employs our recursive homogeneous encoding.

\begin{figure}
\centering
    \includegraphics[width=0.50\textwidth]{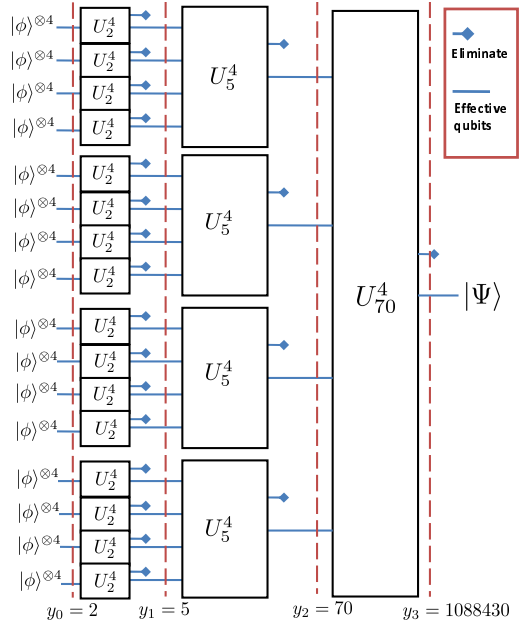}
    \caption{An example of the recursive homogeneous encoding circuit for the
    single-source, $N$-terminal multicast quantum networks. The source generates
    $N$ copies of the quantum state $\ket{\phi}$ in $\cH_d$, and runs the recursive
    homogeneous encoding circuit to output the encoded quantum state $\ket{\Psi}$.
    Let $d=2,k=4,N=64$. There are $\log_k N=3$ recursions in the example. In the
    $n^{th}$ recursion, the encoding unitary $U_{y_{n-1}}^{k}$, where $y_n$ represents
    the dimension of the encoded quantum state after $U_{y_{n-1}}^{k}$, is applied
    to encode each group of group size $k$. The dimension of the input quantum system
    to the last encoding unitary (the encoding complexity) is $D=70^4 \approx 2.4\times10^6$,
    compared to the encoding complexity, $2^{64}\approx 1.84\times10^{19}$, without
    recursive homogeneous encoding. }
\label{tradeoff}
\end{figure}

Denote by $D$ the dimension of the input quantum states to the last encoding
unitary:
\begin{eqnarray}\label{def_D}
D= y^k_{n^*-1}.
\end{eqnarray}
The parameter $D$ captures the complexity of the recursive encoding circuit.
Obviously, the values of $L$ and $D$ depend on the controlling parameter $k$,
which corresponds to the size of each group in our recursive procedure. By
varying $k$ in our recursive procedure, we can provide a trade-off between the
bandwidth requirement $L$ and the encoding complexity $D$. An example of the
recursive homogeneous encoding circuit with $d=2,k=4,N=64$ is illustrated in
Fig.~\ref{tradeoff}.

In the following, we will derive a relation between the minimal bandwidth
requirement $L$ and the size $k$ of each group, and a relation between the
encoding complexity $D$ and the size $k$ of each group in our recursive
procedure. First, following (\ref{recurexact}), we have $k\ll y_n$ when
$n\geq2$.  Then we can approximate the term $y_{n+1}=\h^k_{y_n}$ by:
\begin{eqnarray}\label{approxi}
y_{n+1}=\h^{k}_{y_n}\simeq \frac{{y^k_n}}{k!}=\frac{{y^k_n}}{d^c},
\end{eqnarray}
where we choose a constant $c$ such that $d^c= k!$. Followed from (\ref{def_Ln}), we have
\begin{align}
L_n &=\log_d y_n \nonumber\\
&= \log_d \left(\frac{{y^k_{n-1}}}{d^c}\right) \nonumber\\
&=k\log_d(y_{n-1})-c \nonumber\\
&=kL_{n-1}-c. \label{recursive}
\end{align}
The second equality uses (\ref{recurexact}) and (\ref{approxi}), and the last equality uses (\ref{def_Ln}).
Solving the linear recursive equation (\ref{recursive}) gives
\begin{align}\label{Lnexact}
L_n=k^{n-2}L_2-\frac{k^{n-2}-1}{k-1}c,
\end{align}
where $L_2$ can be derived from (\ref{recurexact}):
$$L_2 = \log_d \frac{(2k)!}{k!k!}.$$
Substituting $N$ for $k^{n^*}$ in (\ref{Lnexact}), we obtain the first desired relation between $L=L_{n^*}$ and $k$:
\begin{eqnarray}\label{final_L}
L = \frac{N}{k^2}L_2-\left(\frac{N}{k^2}-1\right)\frac{c}{k-1}.
\end{eqnarray}
We can obtain the other desired relation between $D$ and $k$ as follows:
\begin{align*}
\log_d D&=k\log_d y_{n^*-1} \\
&=kL_{n^*-1} \\
&=L+c\\
&=\frac{N}{k^2}(L_2-\frac{c}{k-1})+\frac{ck}{k-1}.
\end{align*}
The first equality uses (\ref{def_D}). The second equality uses (\ref{def_Ln}).
The third equality uses (\ref{recursive}). The final equality uses
(\ref{final_L}).

\begin{figure}
\centering
\begin{tabular}{cc}
 \includegraphics[ width=6.3cm , height = 6 cm]{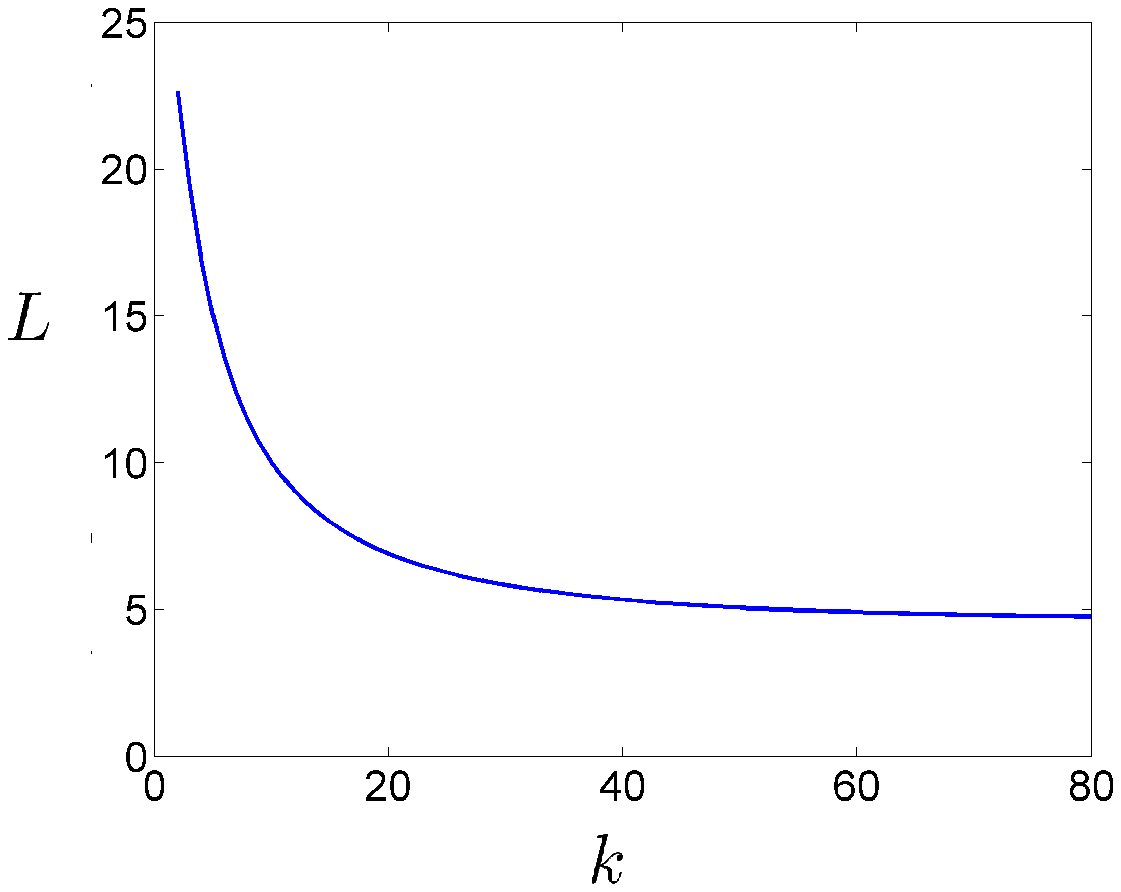}    &
 \includegraphics[ width=6.3cm , height = 6 cm]{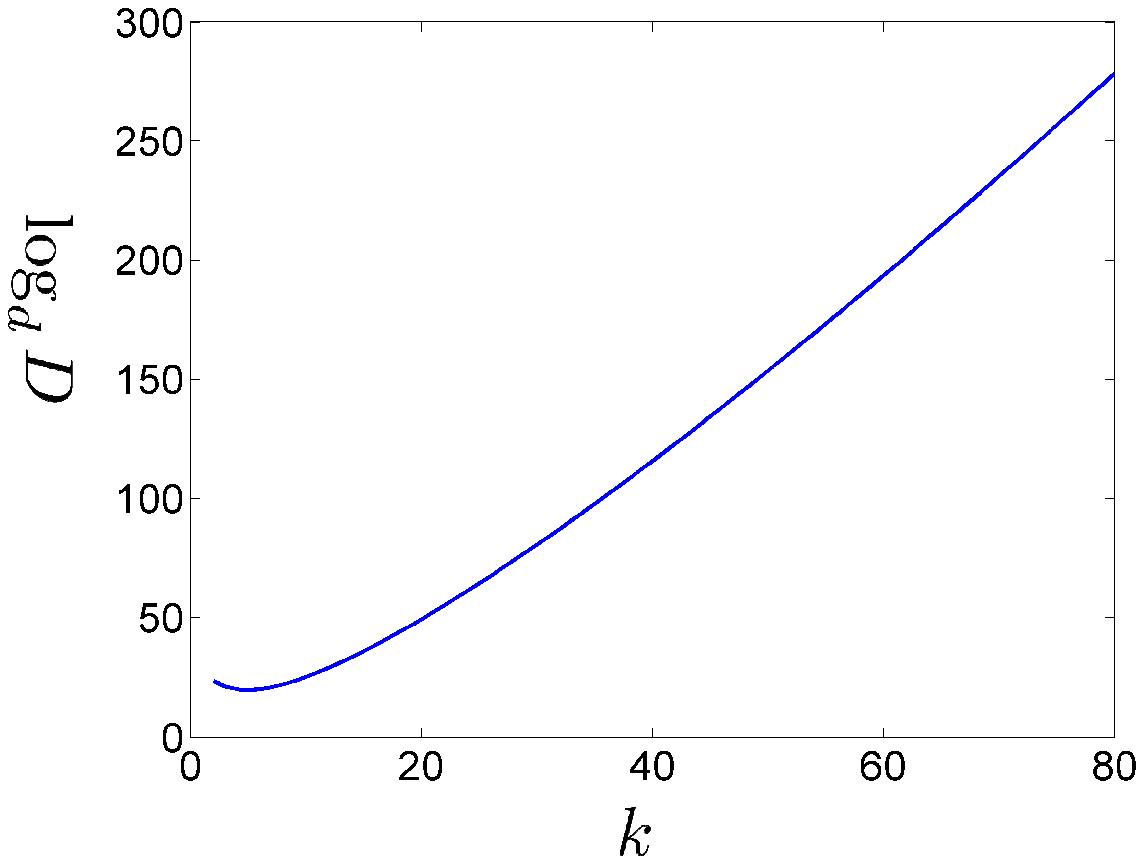}  \\
(a) & (b) \\
\end{tabular}

\caption{(Color online) Figure (a) depicts the numerical evaluation of the
trade-off between the minimal bandwidth requirement $L$ (the number of qudits
per channel use) and the group size $k$ in each recursive procedure. Figure (b)
depicts the numerical evaluation of the trade-off between the encoding
complexity $D$ and the group size $k$ in each recursive procedure. We show
$\log_d D$ in this figure, which corresponds to the number of input qudits to
the final encoding unitary. In both figures, we set $N=80$ and $d=2$. When the
controlling parameter $k$ is small, the bandwidth requirement $L$ is large, but
the encoding complexity is low. When $k$ is large, the bandwidth requirement
$L$ is small, but the encoding complexity $D$ is high.} \label{tradecurve}
\end{figure}

We plot $L$ versus $k$ in Fig.~\ref{tradecurve}(a), and plot $\log_d D$ versus
$k$ in Fig.~\ref{tradecurve}(b). From Fig.~\ref{tradecurve}, we can see that
the group size $k$ in the recursive procedure provides a trade-off between
minimal bandwidth requirement $L$ and the encoding complexity $D$. When $k$ is
small, the bandwidth requirement $L$ is large, but the encoding complexity is
low. When $k$ is large, the bandwidth requirement $L$ is small, but the
encoding complexity $D$ is high. Notice that when $k=N$, we recover Shi and
Soljanin's result in \cite{multicast}.

\section{Heterogeneous encoding in quantum multicast networks}\label{secthetero}
\begin{figure}
\centering
\includegraphics[width=0.56\textwidth]{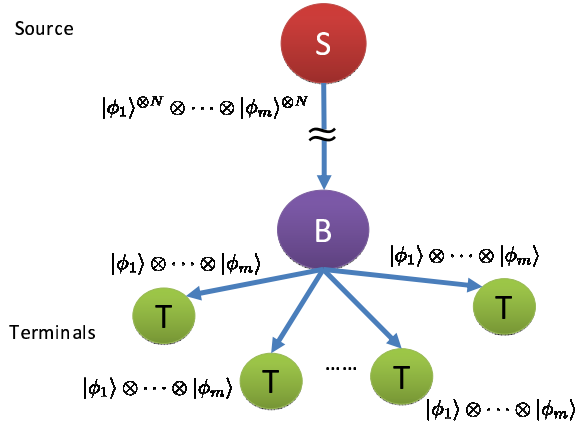}
\caption{(Color online) Quantum multicast networks that contain a single source
(labeled as $S$) and $N$ terminals (labeled as $T$). The source $S$ can generate
$N$ copies of the heterogeneous quantum state $\ket{\Phi}\equiv\ket{\phi_1}\otimes\cdots\otimes\ket{\phi_m}$. The multicasting task is
for $S$ to distribute the quantum state $\ket{\Phi}$ to each terminal $T$ simultaneously
and perfectly through the link that connects the source $S$ and the node $B$.}
    \label{multidata}
    \end{figure}

The main result of this section is the development of two lossless compressing
schemes for the quantum multicast network depicted in Fig.~\ref{multidata}. The
network contains a single source and $N$ terminals. The source $S$ can generate
$N$ copies of the quantum state
$\ket{\Phi}\equiv\ket{\phi_1}\otimes\cdots\otimes\ket{\phi_m}$ in $\cH_{d}^{\otimes
m}$. The multicasting task is for the source $S$ to distribute the quantum
state $\ket{\Phi}$ to each terminal $T$ simultaneously and perfectly through
the link that connects the source $S$ and the node $B$. If no encoding is
performed at the source $S$, the bandwidth (edge capacity) between the source
$S$ and the node $B$ must be as large as $Nm$ qudits for the source to transmit
$\ket{\Phi}^{\otimes N}$ faithfully. We can also apply Shi and Soljanin's
homogeneous encoding \cite{multicast}, or the recursive homogeneous encoding
introduced in the previous section, to encode each $\ket{\phi_i}^{\otimes N}$,
$i=1,2,\cdots,m$.

However, there is a possibility that we can design better lossless compression schemes
than the homogeneous encoding if we know the structure of the set of quantum
states $\{\ket{\phi_1},\cdots,\ket{\phi_m}\}$ generated by the source $S$.  One
such example is when the coefficients of the quantum state $\ket{\phi_i}$ in
the set $\{\ket{\phi_1},\cdots,\ket{\phi_m}\}$ are equal to the
permutation of the coefficients of another quantum state $\ket{\phi_j}$ in the
set. We formally define the set of quantum states whose coefficients are the
same after permutation as follows.
\begin{definition}\label{def_set}
Given is the set of coefficients
$\vec{\alpha}\equiv\{\alpha_0,\alpha_1,\cdots,\alpha_{d-1}\}$, where
$\sum_{i=0}^{d-1}|\alpha_i|^2=1$ and let $|x_0\rangle
\equiv\sum_{i=0}^{d-1}\alpha_i|i\rangle$. We denote by $X(\vec{\alpha})$ the
collection of all the quantum states whose coefficients are the same after all
possible permutations:
\begin{eqnarray}\nonumber
X(\vec{\alpha})\equiv\left\{|\phi\rangle : \forall P,\ |\phi\rangle = P|x_0\rangle\right\},
\end{eqnarray}
where $P$ is an arbitrary $d\times d$ permutation matrix. For example,
$X(\{\alpha,\beta\})=\{\alpha|0\rangle+\beta|1\rangle,\beta|0\rangle+\alpha|1\rangle\}$,
where $|\alpha|^2+|\beta|^2=1$.
\end{definition}
We propose two lossless compression schemes with improvement power for
compressing heterogeneous quantum data $\{\ket{\phi_1},\cdots,\ket{\phi_m}\}$,
if the quantum data generated by the source $S$ is a subset of
$X(\vec{\alpha})$ for a given $\vec{\alpha}$.

Before introducing these two lossless compression schemes, we first show that
there exists a \emph{heterogeneous encoding} for compressing the set of quantum
states $\{\ket{\phi_1},\cdots,\ket{\phi_m}\}\subset X(\vec{\alpha})$. 
Let
$\ket{\Phi}\equiv \ket{\phi_1}\otimes\cdots\otimes\ket{\phi_m}$. It is not
difficult to see that there are $\h^m_d=C^{m+d-1}_{d}$ different coefficients
$\{\gamma_t\}$ in the state $\ket{\Phi}$, and we can represent the state $\ket{\Phi}$ as follows:
\begin{equation}\label{gamma}
\ket{\Phi}=\sum_{t=0}^{\h^m_d-1}\gamma_t\ket{\vec{e}_t},
\end{equation}
where $\ket{\vec{e}_t}$ is similarly defined in (\ref{eq_PI}) and the set $\{\ket{\vec{e}_t}\}$ forms a permutation-invariant basis for $\cH_d^{\otimes
m}$. Then there exists a unitary $S$ such that
\begin{equation}\label{hetero_S}
S\ket{\vec{e}_t}=\ket{0}^{\otimes (m-\log_d\h^m_d)}\otimes\ket{t},
\end{equation}
where $\{\ket{t}\}$ forms an orthonormal basis for $\cH_d^{\otimes \log_d\h^m_d}$. Applying
this unitary $S$ to the heterogeneous quantum state $\ket{\Phi}$ gives
\begin{equation}\label{hetero_cmp}
S\ket{\Phi}=\ket{0}^{\otimes (m-\log_d\h^m_d)}\otimes\left(\sum_{t=0}^{\h^m_d-1}\gamma_t
\ket{t}\right).
\end{equation}
Denote $\ket{\Phi'}=\sum_{t=0}^{\h^m_d-1}\gamma_t \ket{t}$. The unitary $S$
in (\ref{hetero_S}) implements a lossless compression that compresses the
heterogeneous quantum state $\ket{\Phi}$ of dimension $d^m$ to the state
$\ket{\Phi'}$ of dimension $\h_d^m$:
\begin{equation}\label{compress2}
RS\ket{\Phi}=\sum_{t=0}^{\h^m_d-1}\gamma_t \ket{t}=\ket{\Phi'},
\end{equation}
where $R$ is the operation that removes the ancilla state $\ket{0}^{\otimes
(m-\log_d\h^m_d)}$ in (\ref{hetero_cmp}). Furthermore, the compression is
lossless because we can recover the original state $\ket{\Phi}$ from
$\ket{\Phi'}$ as follows:
\begin{equation}\label{decompress2}
S^{-1}A\ket{\Phi'}=\ket{\Phi},
\end{equation}
where $A$ is the operation that adds the ancilla state $\ket{0}^{\otimes
(m-\log_d\h^m_d)}$ to $\ket{\Phi'}$.

In the following, we will propose two new encoding structures that are better
(in terms of bandwidth requirement) than simply encoding each state
$\ket{\phi_i}^{\otimes N}$ separately with the homogeneous encoding when the set
of quantum states $\{\ket{\phi_1},\cdots,\ket{\phi_m}\}$ generated by the
source is a subset of $X(\vec{\alpha})$ for a given coefficient set
$\vec{\alpha}$.

\subsection{Homo-hetero encoding}\label{sec_homo_hetero}
\begin{figure}
\centering
 \includegraphics[width=0.7\textwidth]{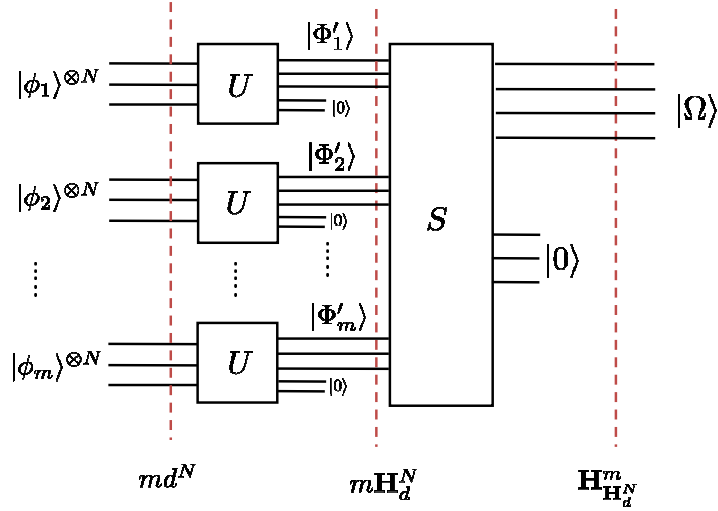}
\caption{(Color online) Quantum circuit for the homo-hetero encoding. The homo-hetero encoding first
uses the homogeneous encoding unitary $U$ to encode each $\ket{\phi_i}^{\otimes N}$ of dimension $d^N$
into an encoded state $\ket{\Phi_i'}$. The dimension of each encoded state $\ket{\Phi_i'}$
is $\h^N_d$. Then the heterogeneous encoding unitary $S$ is applied to encode
$\ket{\Phi_1'}\otimes\cdots\otimes\ket{\Phi_m'}$ into the final state $\ket{\Omega}$ of dimension $\h^m_{\h^N_d}$.
The circuit implements a lossless compression that compresses the
state $\ket{\phi_1}^{\otimes N}\otimes\cdots\otimes\ket{\phi_m}^{\otimes N}$ of dimension $md^N$ to the
encoded state $\ket{\Omega}$ of dimension $\h^m_{\h^N_d}$.}
\label{homofirst_encode}
\end{figure}

The first method for multicasting $\{\ket{\phi_1},\cdots,\ket{\phi_m}\}\subset
X(\vec{\alpha})$ is to first use the homogeneous encoding reviewed in
$(\ref{singleencode1})$ separately on each
$\ket{\Phi_i}\equiv|\phi_i\rangle^{\otimes N}$ to output an encoded state
$|\Phi_i'\rangle$, $\forall i$. It is not difficult to see that the number of
different coefficients in each $\ket{\Phi_i}$ is equal to $\h^N_d$. Denote by
$\vec{\beta}_i$ the collection of all possible coefficients in $\ket{\Phi_i}$.
Each set $\vec{\beta}_i$ is the same because
$\{\ket{\phi_1},\cdots,\ket{\phi_m}\}\subset X(\vec{\alpha})$. Therefore we can
remove the subscript and denote by $\vec{\beta}$. The set of quantum states
$\{\ket{\Phi_1'},\cdots,\ket{\Phi_m'}\}$ is a subset of $X(\vec{\beta})$, since
each state $\ket{\Phi_i'}$ shares the same set of coefficients as the state
$\ket{\Phi_i}$. This allows us to encode
$\ket{\Phi_1'}\otimes\cdots\otimes\ket{\Phi_m'}$ by the heterogeneous encoding
introduced in (\ref{hetero_S}) to an encoded output state $\ket{\Omega}$ of
dimension $\h^m_{\h^N_d}$. We illustrate the above home-hetero encoding in
Fig.~\ref{homofirst_encode}.

The encoding process can be represented by the following expression:
\begin{eqnarray}
|\Omega\rangle = RS\left((RU|\Phi_1\rangle)\otimes\cdots\otimes(RU|\Phi_m\rangle)\right),
\end{eqnarray}
where $\ket{\Phi_i}\equiv\ket{\phi_i}^{\otimes N}$, $U$ and $S$ are the
homogeneous encoding (\ref{singleencode1}) and the heterogeneous encoding
(\ref{hetero_S}), respectively, and $R$ is the operation that removes the
redundant ancilla states.  

The homo-hetero encoding implements a lossless compression that compresses the
state $\ket{\Phi_1}\otimes\cdots\otimes\ket{\Phi_m}$ of dimension $md^N$ to the
encoded state $\ket{\Omega}$ of dimension $\h^m_{\h^N_d}$. This homo-hetero
encoding is lossless because we can recover the original state
$\ket{\Phi_1}\otimes\cdots\otimes\ket{\Phi_m}$ as follows :
\begin{align*}
\ket{\Phi_1'}\otimes\cdots\otimes\ket{\Phi_m'} &= S^{-1}A|\Omega\rangle  \\
\ket{\Phi_1}\otimes\cdots\otimes\ket{\Phi_m} &= (U^{-1}A\ket{\Phi_1'})\otimes\cdots\otimes (U^{-1}A\ket{\Phi_m'}),
\end{align*}
where $A$ is the operator that adds the necessary ancilla states. The node $B$
in Fig.~\ref{multidata} first adds enough amount of the ancilla states to the
state $\ket{\Omega}$. Next he can recover the states $
\ket{\Phi_1'}\otimes\cdots\otimes\ket{\Phi_m'}$ by performing the inverse
heterogeneous encoding $S^{-1}$. Then he adds enough number of the ancilla
states to each state $\ket{\Phi_i'}$, and performs the inverse homogeneous
encoding $U^{-1}$ to generate $N$ copies of the original data $\ket{\phi_i}$,
$\forall i$. Finally he distributes the quantum states to the terminals.

\subsection{Hetero-homo encoding}\label{hetero-homo}
\begin{figure}
\centering
\includegraphics[width=0.55\textwidth]{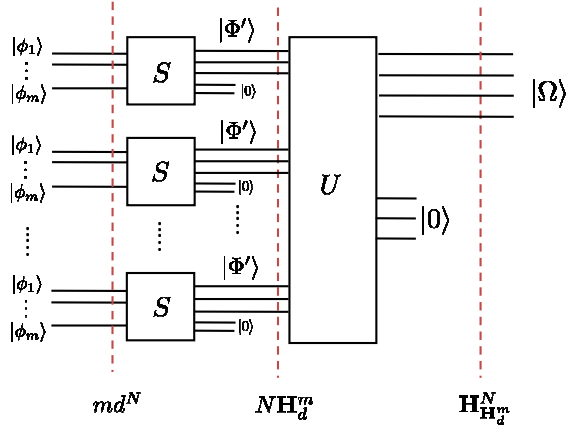}
\caption{(Color online) Quantum circuit for the hetero-homo encoding. The hetero-homo encoding first
uses the heterogeneous encoding unitary $S$ to encode each $\ket{\Phi}\equiv\ket{\phi_1}\otimes\cdots\otimes\ket{\phi_m}$
into an encoded state $\ket{\Phi'}$. The dimension of the encoded state $\ket{\Phi'}$
is $\h^m_d$. Then the homogeneous encoding unitary $U$ is applied to encode
$\ket{\Phi'}^{\otimes N}$ into the final state $\ket{\Omega}$ of dimension $\h^N_{\h^m_d}$.
The circuit implements a lossless compression that compresses the
state $\ket{\phi_1}^{\otimes N}\otimes\cdots\otimes\ket{\phi_m}^{\otimes N}$ of dimension $md^N$ to the
encoded state $\ket{\Omega}$ of dimension $\h^N_{\h^m_d}$.}
\label{heterofirst_encode}
\end{figure}

The second method for multicasting $\{\ket{\phi_1},\cdots,\ket{\phi_m}\}\subset
X(\vec{\alpha})$ is to first apply the heterogeneous encoding $S$
(\ref{hetero_S}) to the state
$\ket{\Phi}\equiv\ket{\phi_1}\otimes\cdots\otimes\ket{\phi_m}$, and outputs the
encoded state $\ket{\Phi'}$ of dimension $\h^m_d$. Next, the homogeneous
encoding $U$ (\ref{singleencode1}) is applied to encode $\ket{\Phi'}^{\otimes
N}$ of dimension $(\h^m_d)^N$ to an encoded state $\ket{\Omega}$ of dimension
$\h^N_{\h^m_d}$. We illustrate the above home-hetero encoding in
Fig.~\ref{heterofirst_encode}.

The hetero-homo encoding process can be expressed as:
\begin{eqnarray}
|\Omega\rangle = RU\left(RS(\ket{\phi_i}\otimes\cdots\otimes\ket{\phi_m})\right)^{\otimes N},
\end{eqnarray}
where $U$ and $S$ are the homogeneous encoding (\ref{singleencode1}) and the
heterogeneous encoding (\ref{hetero_S}), respectively, and $R$ is the operation
that removes the redundant ancilla states.

The hetero-homo encoding implements a lossless compression that compresses the
state $\ket{\Phi}^{\otimes N}$ of dimension $md^N$ to the encoded state
$\ket{\Omega}$ of dimension $\h^N_{\h^m_d}$. This hetero-homo encoding is
lossless because we can recover the original state $\ket{\Phi}^{\otimes N}$ as
follows:
\begin{align*}
\ket{\Phi'}^{\otimes N} &= U^{-1}A|\Omega\rangle  \\
\ket{\Phi}^{\otimes N} &= (S^{-1}A\ket{\Phi'})^{\otimes N},
\end{align*}
where $A$ is the operation that adds the necessary ancilla states. The node $B$
in Fig.~\ref{multidata} first adds enough amount of the ancilla states to the
state $\ket{\Omega}$. Next he can recover the states $ \ket{\Phi'}$ by
performing the inverse homogeneous encoding $U^{-1}$. Then he adds enough
number of the ancilla states to the state $\ket{\Phi'}$, and performs the
inverse heterogeneous encoding $S^{-1}$ to generate $N$ copies of the original
data $\ket{\Phi}$. Finally he distributes the quantum states to the terminals.

\subsection{Comparison and analysis}\label{sectcom}

Table~\ref{comparsiontab} lists the minimal bandwidth requirement of different
encoding methods employed by the source in quantum multicast networks in
Fig.~\ref{multidata}. The quantum multicast networks contain a single source
and $N$ terminals. The set of quantum states
$\{\ket{\phi_1},\cdots,\ket{\phi_m}\}$, $\forall i$ $\ket{\phi_i}$ in $\cH_d$,
to be multicast to each terminal is heterogeneous and is a subset of
$X(\vec{\alpha})$ for a given $\vec{\alpha}$. The source can perform either one
of the following four encoding techniques: multicasting simultaneously without
encoding, the homogeneous encoding (\ref{singleencode1}), the homo-hetero encoding
introduced in section~\ref{sec_homo_hetero}, and the hetero-homo encoding
introduced in section~\ref{hetero-homo}.

\begin{table}
\centering
\begin{tabular}{|l|c|}
  \hline
  Types of Multicasting & Bandwidth Requirement\\
  \hline\hline
  Multicasting directly & $Nm$ \\
  \hline
  Homogeneous encoding & $m\log_d\h^N_{d}$ \\
  \hline
  Homo-hetero encoding  & $ \log_d\h^m_{\h^N_{d}}$ \\
  \hline
  Hetero-homo encoding & $ \log_d\h^N_{\h^m_{d}}$ \\
  \hline
\end{tabular}
\caption{Comparison of minimal bandwidth requirements between the source $S$ and the node $B$ in quantum multicast
networks in Fig.~\ref{multidata}. The set of quantum states to be multicast to
each terminal is of size $m$, and is a subset of
$X(\vec{\alpha})$ for a given set of coefficients
$\vec{\alpha}=\{\alpha_i\}_{i=0}^{d-1}$. The source can perform either one of
the following four encoding techniques: multicasting simultaneously without
encoding, the homogeneous encoding (\ref{singleencode1}), the homo-hetero
encoding introduced in section~\ref{sec_homo_hetero}, and the hetero-homo
encoding introduced in section~\ref{hetero-homo}.} \label{comparsiontab}
\end{table}

\begin{figure}
\centering

 \includegraphics[ width=6cm , height = 5 cm]{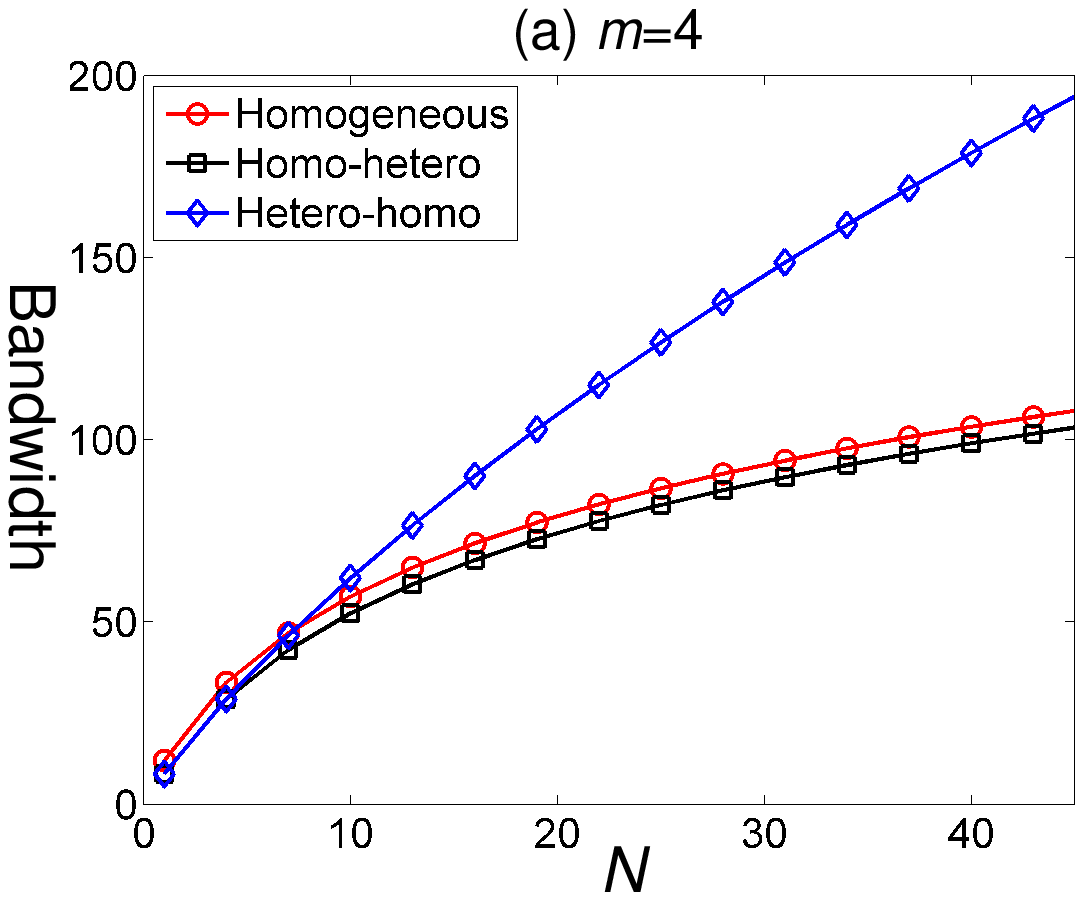}
 \includegraphics[ width=6cm , height = 5 cm]{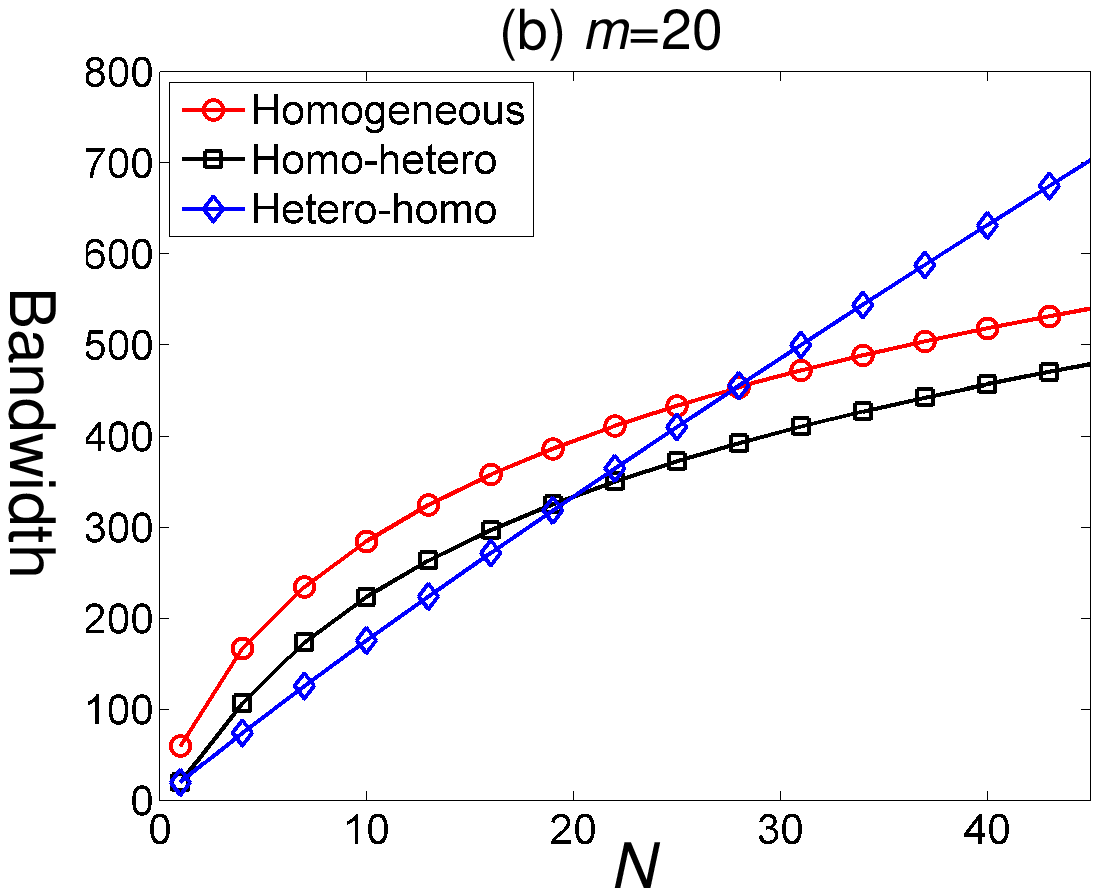}

\caption{(Color online) Numerical evaluation of the minimal bandwidth
requirements of the three non-trivial encoding schemes: the homogeneous encoding
(\ref{singleencode1}), the homo-hetero encoding introduced in
section~\ref{sec_homo_hetero}, and the hetero-homo encoding introduced in
section~\ref{hetero-homo}. The vertical axis represents the minimal bandwidth
requirement (the number of qudits per channel use, $d=8$). The horizontal axis
represents the number $N$ of terminals in quantum multicast networks. We
investigate how the size $m$ of the set of heterogeneous quantum states and the
number $N$ of terminals affect the minimal bandwidth requirements of different
encoding schemes. Figure (a) corresponds to $m=3$, and Figure (b) corresponds
to $m=20$.} \label{comparisionfig}
\end{figure}

We numerically evaluate the minimal bandwidth requirements of the three
non-trivial encoding schemes in Fig.~\ref{comparisionfig}. We investigate how
the size $m$ of the set of heterogeneous quantum states and the number $N$ of
terminals affect the minimal bandwidth requirements of different encoding
schemes. We have the following observations. First, our homo-hetero encoding is
always better than the homogeneous encoding because we perform an extra
heterogeneous encoding (\ref{hetero_S}) to take advantage of the structure of
the encoded quantum states of the homogeneous encoding. Second, the
heterogeneous encoding schemes show obvious gain over the homogeneous encoding
when the size $m$ of the set of heterogeneous quantum states is large (see
Fig.~\ref{comparisionfig}(b)). This is because the compressing power of the
heterogeneous encoding becomes evident when there exists abundant redundancy in
the heterogeneous quantum states. Fig.~\ref{comparisionfig}(b) shows that the
hetero-homo encoding outperforms the other two encodings when $N$ is smaller
than $m$, while the homo-hetero encoding takes the lead when $N$ is larger than
$m$. The reason is the following. When $N$ is large, the redundancy arises
mainly from each quantum state $\ket{\phi_i}^{\otimes N}$, $\forall i$.
Therefore, the first homogeneous compression of the homo-hetero encoding can
remove more redundancy than the first heterogeneous compression of the
hetero-homo encoding. On the other hand, when $N$ is small, the redundancy
mainly comes from the set of heterogeneous quantum states. Hence, the first
heterogeneous compression of the hetero-homo encoding can remove the redundancy
more efficiently than the first homogeneous compression of the homo-hetero
encoding. It also justifies that the cross point of the minimal bandwidth
requirements of these two heterogeneous encodings occurs when $m = N$.

\section{Other quantum multicast networks}\label{sectappli}
In this section, we will consider the quantum multicast networks other than the
single-source, $N$-terminal model depicted in Fig.~\ref{multidata}. Two
examples are discussed. One is the $m$-source, $N$-terminal quantum
multicast networks, and the other is the quantum multicast networks with
ring topology. We then analyze the bandwidth requirements of these two quantum multicast networks, 
where different encoding schemes are employed.

\subsection{Multi-source Multi-terminal networks}
\begin{figure}
\centering
\includegraphics[width=0.55\textwidth]{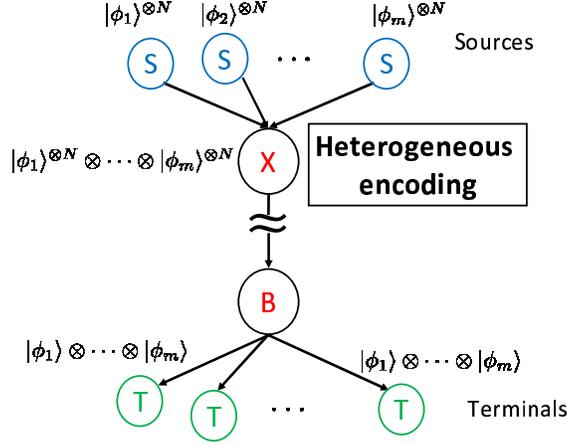}
\caption{(Color online) Quantum multicast networks that contain $m$ sources
(labeled as $S$) and $N$ terminals (labeled as $T$). The $i^{th}$ source $S$ can generate
$N$ copies of a quantum state $\ket{\phi_i}$, $i=1,2,\cdots,m$. The multicasting task is
for each of the sources to distribute his own quantum state, say $\ket{\phi_i}$, to each terminal
$T$ simultaneously and perfectly through the link that connects the node $X$ and the node $B$.
If the set of quantum states $\{\ket{\phi_1},\cdots,\ket{\phi_m}\}$ generated by the sources is a
subset of $X(\vec{\alpha})$ for a given $\vec{\alpha}$, the heterogeneous encoding can be applied
at the node $X$ to efficiently compress the quantum states. }
\label{multisource1}
\end{figure}

One generalization of the single-source, $N$-terminal multicast model is the
$m$-source, $N$-terminal multicast model depicted in Fig.~\ref{multisource1}.
The $i^{th}$ source $S$ can generate $N$ copies of a quantum state
$\ket{\phi_i}$ in $\cH_d$, $i=1,2,\cdots,m$, such that the set of quantum
states $\{\ket{\phi_1},\cdots,\ket{\phi_m}\}$ is a subset of $X(\vec{\alpha})$
for a given $\vec{\alpha}$. The multicasting task is for each of the sources to
distribute his own quantum state, say $\ket{\phi_i}$, $i=1,2,\cdots,m$, to each terminal
simultaneously and perfectly through the link that connects the node $X$ and
the node $B$. If no encoding is performed at the node $X$, the bandwidth
required between the node $X$ and the node $B$ must be as large as $Nm$ qudits
per channel use. The node $X$ can also perform either the homogeneous or the heterogeneous
encoding to compress the quantum states. The minimal bandwidth
requirements of the link connecting the node $X$ and the node $B$ are
depicted in Table~\ref{comparsiontab}, depending
on the encoding technique employed by the node $X$.

\subsection{Ring networks}
\begin{figure}
\centering
\includegraphics[width=0.55\textwidth]{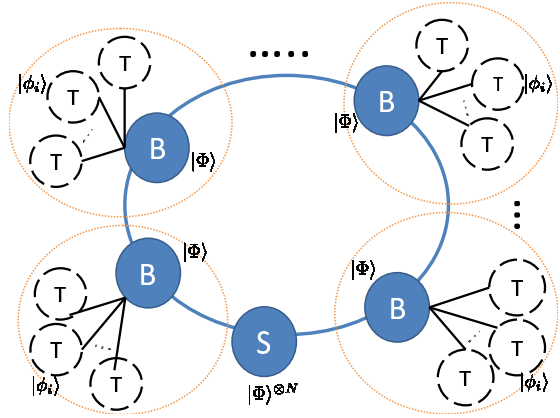}
\caption{(Color online) The quantum multicast networks with ring topology that contain a
single-source (labeled by $S$), and $N$ nodes (labeled by $B$) on a ring. Each node $B$
connects to $m$ terminals (labeled by $T$), and forms a cluster. The source $S$ can
generate $N$ copies of the quantum state $\ket{\Phi}\equiv\ket{\phi_1}\otimes\cdots\otimes\ket{\phi_m}$.
The multitasking task is for the source $S$
to perfectly distribute the $i^{th}$ quantum state in $\{\ket{\phi_1},\cdots,\ket{\phi_m}\}$ to the $i^{th}$ terminal
in each of the $N$ clusters, $\forall i$.}
\label{multisource2}
\end{figure}

We consider the quantum ring networks depicted in Fig.~\ref{multisource2}. The
networks contain a single-source $S$ and $N$ number of nodes on a ring. Each
node $B$ connects to $m$ number of terminals, and forms a cluster. We consider the following multitasking task with the
assumption that $N\gg m$: the source $S$ would like to simultaneously and
perfectly distribute the $i^{th}$ quantum state in $\{\ket{\phi_1},\cdots,\ket{\phi_m}\}$ to the
$i^{th}$ terminal in each of the $N$ clusters, $\forall i$. The source $S$ can use the
clockwise and counterclockwise paths to distribute the quantum states to
destinations. We will evaluate the overall bandwidth required on the ring in
order for $S$ to accomplish the desired multicasting task.

If no encoding is applied, the source simply sends $\frac{N}{2}$ copies of the
quantum state $\ket{\Phi}$, where
$\ket{\Phi}\equiv\ket{\phi_1}\otimes\cdots\otimes\ket{\phi_m}$, down to both
the clockwise and counterclockwise paths. The first node on both sides
distributes one copy of the state $\ket{\Phi}$ to the $m$ terminals in his
cluster, and forwards the rest $\frac{N}{2}-1$ copies of the state $\ket{\Phi}$
to the next node. The process continues until all the terminals receive a
desired quantum state. We can then evaluate the overall bandwidth required on
the ring as follows:
\begin{eqnarray}
2m\sum^{\frac{N}{2}}_{k=1}k = m \frac{N}{2}(\frac{N}{2}+1).
\end{eqnarray}
The bandwidth consumption without encoding is $O(N^2)$.

If the source and the $N$ number of nodes on the ring employ the homogeneous
encoding, the multicasting task proceeds as follows. For each $i=1,2,\cdots,m$, the source $S$ applies two
instances of the homogeneous encoding on $\frac{N}{2}$ copies of the state $\ket{\phi_i}$, and then
sends the encoded states down to both
paths. The dimension of the encoded quantum state on either path is
$m\h^{\frac{N}{2}}_d$. After receiving the encoded quantum state, the first
node on both sides performs the inverse of the homogeneous encoding to recover
the original state $\ket{\Phi}^{\otimes \frac{N}{2}}$, distributes one copy of
the state $\ket{\Phi}$ to the terminals in his cluster, and then applies the
homogeneous encoding again to encode the rest $\frac{N}{2}-1$ copies of the
state $\ket{\Phi}$. The dimension of the encoded state now becomes
$m\h^{\frac{N}{2}-1}_d$.  The process continues until all the terminals receive
a desired quantum state. We can then evaluate the overall bandwidth required on
the ring when each node applies the homogeneous encoding as follows:
\begin{eqnarray}\label{Bhomo}
2m\sum^{\frac{N}{2}}_{k=1}\log_d\h^k_{d} &=& 2m \sum_{k=1}^{\frac{N}{2}}\log_d \frac{(k+d-1)!}{d!(k-1)!} \nonumber \\
&=& 2m \sum_{\ell=0}^{d-1} \sum_{k=1}^{\frac{N}{2}} \log_d (k+\ell) - mN \log_d d! \nonumber \\
&\simeq& 2md \sum^{\frac{N}{2}}_{k=1}\log_d k - mN \log_d d! \label{eq_first}\\
&\simeq &  mN d\log_d \frac{N}{2}. \label{eq_second}
\end{eqnarray}
The first approximation follows from the assumption $N\gg d$.
The second approximation holds because $mN \log_d d!$ is much smaller the first term in (\ref{eq_first}) and it can be evaluated by the following:
\begin{eqnarray*}
\sum^{N/2}_{k=1}\log_d k \simeq \int^{N/2}_{1}\log_d kdk = \frac{1}{\ln d}\left(k\ln k-k\right)\mid^{\frac{N}{2}}_{1}.
\end{eqnarray*}

If the source and the $N$ number of nodes on the ring employ the heterogeneous
encoding, the multicasting task proceeds as follows. Specifically, we will use the
homo-hetero encoding because it outperforms the hetero-homo encoding when $N\gg
m$. The source $S$ applies two instances of the
homo-hetero encoding on $\frac{N}{2}$ copies of the state $\ket{\Phi}$,
and then sends the two encoded states down to both paths. The
dimension of the encoded quantum state in either path is
$\h^m_{\h^{\frac{N}{2}}_d}$. After receiving the encoded quantum state, the
first node on both sides performs the inverse of the homo-hetero encoding to
recover the original state $\ket{\Phi}^{\otimes \frac{N}{2}}$, distributes one
copy of the state $\ket{\Phi}$ to the terminals in his cluster, and then
applies the homo-hetero encoding again to encode the rest $\frac{N}{2}-1$ copies
of the state $\ket{\Phi}$. The dimension of the encoded state now becomes
$\h^m_{\h^{\frac{N}{2}-1}_d}$.  The process continues until all the terminals
receive a desired quantum state. We can then evaluate the overall bandwidth
required on the ring when each node applies the homo-hetero encoding as
follows:
\begin{eqnarray}
2\sum_{k=1}^{\frac{N}{2}}\log_d \h^m_{\h^{k}_d}&=& 2 \sum_{k=1}^{\frac{N}{2}}\log_d \frac{\left(m+\h^k_d -1\right)!}{(m-1)! \h^k_d!}\nonumber\\
&=& 2\sum_{\ell=1}^{m-1}\sum_{k=1}^{\frac{N}{2}}\log_d \left(\h^k_d+\ell\right) -N \log_d (m-1)!\nonumber\\
&\simeq& 2(m-1)\sum_{k=1}^{\frac{N}{2}}\log_d\h^k_d-N\log_d (m-1)! \\
&\simeq& (m-1)N d \log_d \frac{N}{2}. \label{eq_third}
\end{eqnarray}
The first approximation holds because $\h^k_d$ is usually very large. The second approximation follows from (\ref{eq_second}):
$$\sum_{k=1}^{\frac{N}{2}}\log_d\h^k_d\simeq \frac{N}{2}d \log_d \frac{N}{2}.$$

The overall bandwidth consumed by both the homogeneous and heterogeneous encoding is $O(N\log N)$. The bandwidth saved due to encoding is
on the order of $O(\frac{1}{N}\log N)$.

\section{Conclusion}\label{sectdiscuss}

The achievement of this paper is two-fold. First, we proposed a novel recursive
homogeneous encoding to realize quantum multicasting with low encoder
complexity in section~\ref{secthomo}. Our recursive homogeneous encoding
circuit can provide a reasonable trade-off between the encoder complexity and
the dimension of the encoded state (corresponds to the bandwidth requirement of
the quantum networks). Though the encoding proposed by Shi and Soljanin
\cite{multicast} reduced the minimal bandwidth requirement from $N$  to $\log_d
\h^N_d$, the hardware complexity of their encoding circuit is daunting. Hence, it is
difficult to practically implement their encoder in quantum multicast networks. 
Our recursive encoding idea proves to be
extremely useful in the situation where the technology of producing large-scale 
quantum circuits is limited. We detailed the relation between the minimal
bandwidth requirement and the encoding complexity. One can easily decide the
dimension of the compressed state and the encoder complexity by our formula. We
also analyzed how the relation is affected by the size $k$ of the divided group
in each recursion.

The second achievement of this paper is the proposal of the heterogeneous
encodings that further improve the compressing power of Shi and Soljanin's
encoding when the set of quantum states satisfies the condition in
definition~\ref{def_set}. When the size $m$ of the heterogeneous quantum states
is larger than the number $N$ of destinations, the hetero-homo encoding is the
most efficient. On the other hand, when $N>m$, the homo-hetero encoding
outperforms the other encoding schemes. The heterogeneous encoding can be
applied in several quantum multicast networks, including the single-source, $N$
terminal model, the multi-source  multi-terminal model, and the ring networks.
The bandwidth requirements for these network models are analyzed.

We can implement a recursive heterogeneous encoding if we wish to reduce the
complexity of the heterogeneous encoding. The implementation is similar to the
recursive homogeneous encoding proposed in section~\ref{secthomo}. Since both of
the homogeneous encoding and the heterogeneous encoding are lossless compression,
we believe the recursive version of these compression schemes will find its
applications in many other areas.

\bibliographystyle{IEEEtran}
\bibliography{mybib}

\begin{thebibliography}{10}
\providecommand{\url}[1]{#1}
\csname url@samestyle\endcsname
\providecommand{\newblock}{\relax}
\providecommand{\bibinfo}[2]{#2}
\providecommand{\BIBentrySTDinterwordspacing}{\spaceskip=0pt\relax}
\providecommand{\BIBentryALTinterwordstretchfactor}{4}
\providecommand{\BIBentryALTinterwordspacing}{\spaceskip=\fontdimen2\font plus
\BIBentryALTinterwordstretchfactor\fontdimen3\font minus
  \fontdimen4\font\relax}
\providecommand{\BIBforeignlanguage}[2]{{%
\expandafter\ifx\csname l@#1\endcsname\relax
\typeout{** WARNING: IEEEtran.bst: No hyphenation pattern has been}%
\typeout{** loaded for the language `#1'. Using the pattern for}%
\typeout{** the default language instead.}%
\else
\language=\csname l@#1\endcsname
\fi
#2}}
\providecommand{\BIBdecl}{\relax}
\BIBdecl

\bibitem{Ahlswede00networkinformation}
R.~Ahlswede, N.~Cai, S.-Y.~R. Li, and R.~W. Yeung, ``Network information
  flow,'' \emph{IEEE Trans. Inform. Theory}, vol.~46, pp. 1204--1216, 2000.

\bibitem{Ho06arandom}
T.~Ho, M.~M\'edard, R.~Koetter, D.~R. Karger, M.~Effros, J.~Shi, and B.~Leong,
  ``A random linear network coding approach to multicast,'' \emph{IEEE Trans.
  Inform. Theory}, vol.~52, pp. 4413--4430, 2006.

\bibitem{citeulike:1181440}
R.~W. Yeung, S.-Y. Li, and N.~Cai, \emph{Network Coding Theory}.\hskip 1em plus
  0.5em minus 0.4em\relax {Now Publishers}, June 2006.

\bibitem{KGNR09}
H.~Kobayashi, F.~Le~Gall, H.~Nishimura, and M.~Roetteler, ``General scheme for
  perfect quantum network coding with free classical communication,''
  \emph{Automata, Languages and Programming\emph{, Springer, LNCS}}, vol. 5555,
  pp. 622--633, 2009.

\bibitem{noclone}
W.~K. Wootters and W.~H. Zurek, ``A single quantum cannot be cloned,''
  \emph{Nature}, vol. 299, pp. 802 -- 803, Oct. 1982.

\bibitem{nobroadcast}
H.~{Barnum}, C.~M. {Caves}, C.~A. {Fuchs}, R.~{Jozsa}, and B.~{Schumacher},
  ``Noncommuting mixed states cannot be broadcast,'' \emph{Phys. Rev. Lett.},
  vol.~76, pp. 2818--2821, Apr. 1996.

\bibitem{HINRY07}
M.~Hayashi, K.~Iwama, H.~Nishimura, R.~Raymond, and S.~Yamashita, ``Quantum
  network coding,'' \emph{Computer Science\emph{, Springer, LNCS}}, vol. 4393,
  pp. 610--621, 2007.

\bibitem{butterfly}
D.~Leung, J.~Oppenheim, and A.~Winter, ``Quantum network communication -- the
  butterfly and beyond,'' 2006, quant-ph/0608223.

\bibitem{Hayashi07}
M.~Hayashi, ``Prior entanglement between senders enables perfect quantum
  network coding with modification,'' \emph{Physical Review A}, vol.~76, p.
  040301, 2007.

\bibitem{multicast}
Y.~Shi and E.~Soljanin, ``On multicast in quantum networks,'' in
  \emph{Proceedings of the 40th Annual Conference on Information Sciences and
  Systems}, Mar. 2006, pp. 871--876.

\bibitem{distributed}
K.~Van~Meter, R.~Nemoto and W.~Munro, ``Communication links for distributed
  quantum computation,'' \emph{IEEE Transactions on Computers}, vol.~56,
  no.~12, pp. 1643--1653, Dec. 2007.

\bibitem{wireless}
S.-T. Cheng, C.-Y. Wang, and M.-H. Tao, ``Quantum communication for wireless
  wide-area networks,'' \emph{IEEE Journal on Selected Areas in
  Communications}, vol.~23, no.~7, pp. 1424--1432, Jul. 2005.

\bibitem{multipartyGHZ}
H.~Ma and B.~Chen, ``Quantum network based on multiparty quantum secret
  sharing,'' \emph{Eighth ACIS International Conference on Software
  Engineering, Artificial Intelligence, Networking, and Parallel/Distributed
  Computing}, vol.~2, pp. 347--351, Aug. 2007.

\bibitem{multipartyEPR}
F.-G. Deng, X.-H. Li, C.-Y. Li, P.~Zhou, and H.-Y. Zhou, ``Multiparty quantum
  state sharing of an arbitrary two-particle state with
  {E}instein-{P}odolsky-{R}osen pairs,'' \emph{Phys. Rev. A.}, vol.~72, p.
  044301, 2005.

\bibitem{NC00}
M.~A. Nielsen and I.~L. Chuang, \emph{Quantum Computation and Quantum
  Information}.\hskip 1em plus 0.5em minus 0.4em\relax New York: Cambridge
  University Press, 2000.

\bibitem{CT91}
T.~M. Cover and J.~A. Thomas, \emph{Elements of Information Theory}, ser.
  Series in Telecommunication.\hskip 1em plus 0.5em minus 0.4em\relax New York:
  John Wiley and Sons, 1991.

\end{thebibliography}

\end{document}